# Predicting First-Year Dropout from Pre-Enrolment Motivation Statements Using Text Mining

Karlijn F. B. Soppe*[#], Ayoub Bagheri*, Shiva Nadi[+], Irene G. Klugkist*, Theo Wubbels[¥], & Leoniek D.N.V. Wijngaards-de Meij*

[*]Department of Methods and Statistics, Faculty of Social and Behavioral Sciences, Utrecht University

[+]Direction of Information and Technology Services, Utrecht University

[¥]Department of Education, Faculty of Social and Behavioral Sciences, Utrecht University

[#]corresponding author email: k.f.b.soppe@uu.nl

**Abstract**

Preventing student dropout is a major challenge in higher education and it is difficult to predict prior to enrollment which students are likely to drop out and which students are likely to succeed. High School GPA (HSGPA) is a strong predictor of dropout, but much variance in dropout remains to be explained. This study focused on predicting university dropout by using text mining techniques with the aim of exhuming information contained in students' written motivation. By combining text data with classic predictors of dropout (student characteristics), we attempt to enhance the available set of predictive student characteristics. Our dataset consisted of 7,060 motivation statements of students enrolling in a non-selective bachelor at a Dutch university in 2014 and 2015. Support Vector Machines (SVMs) were trained on 75% of the data and several models were estimated on the test data (25%). We used various combinations of student characteristics (e.g., high school grades, age) and text (i.e., TFiDF, topic modelling, LIWC dictionary). Results showed that, although the combination of text and student characteristics did not improve the prediction of dropout, text analysis alone predicted dropout similarly well as a set of student characteristics. Suggestions for future research are provided.

# 1. Introduction

Improving student retention is one of the biggest challenges in higher education. Retaining students results in higher revenue for universities (Zhang et al., 2010) since their funding is often partially based on graduation rates (Jongbloed et al., 2018). For students, finalizing their degree is also important, as dropping out of higher education is associated with negative consequences, such as untapped human potential, a low return on their financial investment (Psacharopoulos, 1994), and reduced social welfare (Hällsten, 2017). Moreover, low retention rates also impact society since income levels rise with a higher education degree (Jayaraman, 2020). Thus, it is paramount for society to keep dropout in higher education to a minimum.

Ideally, students at risk of dropout should be identified prior to enrollment, to minimize negative consequences for both students and universities. In selective admission, it is common practice to try to identify students at risk of dropout based on their application. Staff members of the admissions committee are generally looking for both cognitive (e.g., prior performance) and non-cognitive (e.g., personality and motivation) factors when selecting suitable candidates (Kurysheva et al., 2019). The use of some of these non-cognitive criteria, especially by means of motivation and recommendation letters, for selecting students has been subjected to criticism (Kira Talent, 2018; Posselt, 2016). Self-report measures such as motivation letters are susceptible to faking by the applicant, when being used in a high-stakes context (Niessen et al., 2017). Moreover, filtering out true motivation can be challenging for program staff. They may need to "read between the lines" to form an idea about the factors driving a student to apply for their study program. Furthermore, it might be hard to identify students' motivation solely based on a written statement. Characteristics of the reader (e.g., experience), their psychology, and environment can introduce bias into the evaluation of the motivation letters (Bridgeman, 2013). These aspects make humans inconsistent and unreliable evaluators (Zupanc, 2018). Lastly, reading these statements is very time consuming and it is not easy to compare motivation across students. This cumbersomeness may make program staff less likely to engage in evaluating motivation texts as part of the enrollment procedure (Moskal et al., 2016). All things considered, the vast amount of text data in the application process may exceed the human capacity to process it thoroughly.

This study, therefore, focuses on predicting university dropout by using text mining techniques to exhume information contained in students' written motivation. The aim of this study is to investigate whether this novel approach can disclose information present in text, and thereby contribute to detecting students who are potentially at risk of dropout as early as possible. If so, traditional prediction models could be updated, using these techniques, to obtain higher predictive power.

Using machine learning techniques in education is not a new phenomenon (see Foster and Francis, 2020 for a systematic review), but almost all Educational Data Mining (EDM) research on student dropout prediction used structured data (i.e., quantitative, alpha-numeric data that can directly be used as input in statistical models). There are, however, some studies using Natural Language Processing (NLP) techniques and unstructured data (i.e., qualitative data in no particular format, such as text, audio, or video files) in predicting student completion of Massive Open Online Courses (MOOCs). Most of these studies use sentiment analysis to detect positive or negative phrases, motivation, engagement, etc. in discussion forums or assignments (Jayaraman, 2020). For example, in a study on students' opinions towards a course, Wen and colleagues (2014) found, using sentiment analysis, that students who used words related to motivation were more likely to complete the course. Moreover, Crossley and colleagues (2016) used NLP techniques on MOOC forum posts and found that a range of NLP indicators, such as lexical sophistication and writing fluency, were predictive of student completion of the MOOC.

Outside of MOOCs, we know of only two studies that used text mining and NLP techniques to add to the early identification of students at risk of dropout by improving the predictive validity of student dropout models using unstructured data. One of these studies used sentiment analysis to predict dropout by analyzing notes written by student advisors (Jayaraman, 2020). The most frequently used words were "drop" (negative sentiment) and "progress" (positive sentiment). By comparing several models, the most accurate model was sought to predict dropout. The best performing model, a random forest classifier, predicted dropout with 73% accuracy. In the second study, Stone and colleagues (2019) used both human coding and a variety of NLP techniques to detect non-cognitive traits, such as psychological connection (which they used as a proxy for intrinsic motivation), by analyzing students' 150-word open-ended descriptions of their own extracurricular activities or work experiences included in their college applications. Correlations between human coding and model-based coding ranged from medium-small to medium-strong on the respective non-cognitive traits. The correlation between human- and model-based coding for psychological connection was .655, indicating a medium-strong relation. The non-cognitive traits were then used in separate regression models to predict 6-year graduation outcomes. Psychological connection was insignificant in both the human- and the model-coded prediction of 6-year graduation. However, results showed that some other traits had predictive power net of other known predictors. For example, results from both the human- and model-based coding models showed that students portraying a growth mindset were more likely to graduate within six years, when controlling for sociodemographics, secondary school GPA and intelligence.

In this study, having the aim to contribute to early detection of at-risk students, we use NLP-based techniques to analyze short motivation statements of applicants to non-selective bachelor programs (i.e., programs that admit all students who obtained a pre-university diploma) in the Netherlands. In doing so, we try to answer the question *whether students at risk of dropout can be identified through text mining, based on their motivation for the study program as written in their intake questionnaire prior to enrollment; and whether information extracted from these motivation statements adds predictive power net of student characteristics.*

## 2. Methods
### 2.1. Dataset
All students applying to a higher education study program in the Netherlands have to file an application through a central website. Upon filing such an initial request, students applying to a non-selective bachelor will receive a request to participate in a matching procedure. This procedure consists of an online questionnaire regarding motivation and student background characteristics and may be followed by an activity online or on campus (e.g., interview, online course, or matching day). Upon completion of the matching procedure, students will receive a request to finalize their enrollment.

For this study, we obtained 7,060 student records from a Dutch university, composing all students who enrolled in a non-selective bachelor at this university during the academic years 2014 and 2015. The obtained academic records consisted of student background information, their answers to the matching questionnaire, and student progress data such as first-year dropout. The dataset is analyzed using Python and source code can be found on GitHub[1].

### 2.2. Variables
In this study both structured data (i.e., a set of student characteristics ranging from prior education to the number of study programs a student applied for) and unstructured data (i.e., motivation statements) were used to predict first-year student dropout. Below, we discuss the operationalization of the structured data. In the next sections we will elaborate on the unstructured data and how features were extracted from the raw text data.

**Dropout**[2]. Student dropout is measured using information on whether a student re-enrolled in the second year of the study program. Students who paid the tuition fees for the second year are considered to have re-enrolled for their sophomore year. They are treated as the retention class (0). Students who did not pay tuition for their sophomore year are classified as dropouts (1).

---

[1] https://github.com/ShNadi/study_motivation
[2] Whether students paid for their re-enrollment is one way of operationalizing dropout. Another way is to look at whether students have obtained sufficient credits to continue. In the Netherlands, students must obtain 45/60 ECTS to continue in their sophomore year. For this study we ran all models with this classifier (i.e., did or did not obtain 45 ECTS) as well and results differ only marginally.

**Prior education.** Students with a variety of educational backgrounds enroll in Dutch universities. Prior education was measured using a categorical variable with the following levels: preparatory university diploma (VWO), to be obtained (1); preparatory university diploma (VWO), already obtained (2); university of applied sciences propaedeutic diploma (3); and other (4).

**High school performance.** In the Netherlands, high school grades are given on a scale from 1-10, with 10 representing a perfect score. In the questionnaire prospective students were requested to self-report their grades for the three core subjects in high school (Dutch, English, and Mathematics). To measure high school performance, a mean grade of these three core subjects in the pre-final year of high school was calculated. If one of the grades was missing, a mean of the remaining two subjects was taken, otherwise the variable was coded as missing.

**Ability beliefs.** Students' belief in their own abilities was measured using a single item from the questionnaire, stating: "the program matches my capacities and skills". The item could be answered with (1) yes or (0) no.

**Interests.** Students' interest in the study program was measured using the statement "the program is in line with my interests". The item could be answered with (1) yes or (0) no.

**Gender.** Information about students' gender was taken from the university registration systems. Male students were coded (1) and female students as (0).

**Age.** Students' date of birth was taken from the university registration systems and then recoded into age at the start of the study program by subtracting date of birth from the first day of the academic year for each cohort.

**Cohort.** The dataset consists of students who enrolled in a non-selective bachelor's program during the academic years of 2014, coded as (1) and 2015, coded as (2).

**Study program.** In the Netherlands, students choose a specific study program to enroll in for their university education, e.g., Psychology, Chemistry, Law. To control for differences in dropout across study programs, all study programs were added as dichotomous variables, i.e., coded as (1) if a student applied for that program and as (0) otherwise.

**Discipline.** All study programs were allocated into three different disciplines: Science, Technology, Engineering & Mathematics (1), Social Sciences (2) and Humanities (3).

**Previously enrolled.** Students who have previously been enrolled in another study program at the same university were coded (1) and all others as (0).

**Multiple requests.** Students who filed admission requests for more than one study program were coded (1) and all others as (0).

Table 1 provides an overview of the descriptive statistics of the structured data in the sample, split by cohort.

**Table 1.** Descriptive statistics of the structured data per cohort.

| Characteristics | 2014 | | | | 2015 | | | |
|---|---|---|---|---|---|---|---|---|
| | Re-enrolled | | Dropout | | Re-enrolled | | Dropout | |
| | n | M(SD) | n | M(SD) | n | M(SD) | n | M(SD) |
| Prior education | | | | | | | | |
|   pre-university; to be obtained | 1785 | | 458 | | 1619 | | 480 | |
|   pre-university; obtained | 793 | | 293 | | 636 | | 276 | |
|   propaedeutic diploma | 309 | | 96 | | 284 | | 95 | |
|   other | 63 | | 13 | | 54 | | 16 | |
| High school performance | 2843 | 6.84 (.61) | 817 | 6.64 (.56) | 2316 | 6.89 (.66) | 775 | 6.62 (.59) |
| Ability beliefs | | | | | | | | |
|   positive | 2092 | | 616 | | 1801 | | 562 | |
|   negative | 858 | | 244 | | 792 | | 305 | |
| Interests | | | | | | | | |
|   positive | 2895 | | 830 | | 2401 | | 783 | |
|   negative | 55 | | 30 | | 192 | | 84 | |
| Gender | | | | | | | | |
|   male | 1267 | | 455 | | 1184 | | 506 | |
|   female | 1683 | | 405 | | 1409 | | 361 | |
| Age | 2837 | 19.00 (2.44) | 816 | 19.35 (2.43) | 2593 | 18.95 (2.39) | 867 | 19.50 (3.03) |
| Discipline | | | | | | | | |
|   STEM | 677 | | 248 | | 612 | | 233 | |
|   Social sciences | 1739 | | 419 | | 1413 | | 416 | |
|   Humanities | 534 | | 193 | | 568 | | 218 | |
| Previously enrolled | | | | | | | | |
|   yes | 284 | | 93 | | 251 | | 114 | |
|   no | 2666 | | 767 | | 2342 | | 753 | |
| Multiple requests | | | | | | | | |
|   yes | 107 | | 40 | | 107 | | 39 | |
|   no | 2843 | | 820 | | 2486 | | 828 | |

*2.2.1. Preprocessing the motivation statements*

To analyze unstructured text data, several steps need to be taken to reduce its high dimensionality. Our raw text data consist of students' answer on the intake questionnaire to the following question: "Why do you want to study [program name] in [city]? (10-25 lines)". This question was followed by the instruction: "When answering the question, for example think about your motivation for the content of the study program, your choice for an academic program, and your motivation for a profession or position that this program prepares you for". The first step that needs to be taken is pre-processing the data. In our analysis, pre-processing the motivation statements consists of stop word removal, removing whitespaces and numbers, and converting text into lowercases. This is a step that enhances the performance of the algorithm in later stages. After the pre-processing step, the high dimensionality of text data still prevents it from being used directly in a statistical model and therefore a feature engineering step is required to inspect the text data and acquire multiple feature sets.

*2.2.2. Feature engineering*

Feature engineering is a process in machine learning that is used to extract analyzable properties from raw data. In machine learning these analyzable properties are known as features; they can be considered independent variables. In this study, three different types of feature engineering were applied to the motivation statements. Initially, a bag-of-words representation was used as a simplified representation of the text data. Thereafter, two additional, and more advanced, feature engineering methods were applied; latent Dirichlet allocation topic modelling, and linguistic inquiry and word count (LIWC) dictionary words. Each of these methods is explained below.

**Bag-of-Words.** This is a process that converts text data into numbers in a (e.g., document-term) matrix. To create this matrix, we used Term Frequency inverse Document Frequency (TFiDF) which is a bag-of-words method intended to reflect the relative frequency of a term (word) in each document (motivation statement). TFiDF can be calculated by multiplying the number of times a word appears in a document, and the inverse document frequency of the word in the dataset. With TFiDF, the words that are common in every document rank low even though they appear many times. This is because TFiDF is offset by the number of documents that contain the word. The bag-of-words representation is the simplest way to make text analyzable in a statistical model. In the remainder of this paper, we will refer to it as just "text" or the method we used: "TFiDF".

**Topic modeling.** One way of reducing the high dimensionality of text data, is to represent it as a set of topics across documents. So, instead of looking at the word frequency of single words like TFiDF does, words are clustered into groups that represent underlying concepts. To identify topics in the motivation statements we employed a topic modeling method, Latent Dirichlet allocation (LDA), using a collapsed Gibbs sampling approach (Blei et al., 2003). LDA topic modeling considers each topic as a probability distribution over terms, and each document as a combination of topics. LDA is an unsupervised method, which means that it can extract hidden topics from text without human assistance. This entails that the extracted topics do not always construe a clear meaning. Therefore, it is up to the researchers to identify which number of topics is the best conceptual representation of the text data. Therefore, we ran a set of models with different numbers of topics (i.e., 5, 10, 15, 20, 50). Based on inspection of the representative terms in each topic and to what extent these terms could form a meaningful topic together, two of the authors independently selected the model with 15 topics as the best representation of the data. Therefore, the 15 topics were used as "features extracted from text".

**Linguistic Inquiry and Word Count (LIWC).** LIWC is a text analysis tool that reduces the dimensionality of the data by mapping words onto predetermined categories, using psychometrically validated dictionaries. LIWC can expose certain psychological characteristics of the writer (Tausczik & Pennebaker, 2012). The main categories provided by LIWC are general information (e.g., word count), linguistic dimensions (categorized in verbs and function words, such as pronouns), psychological processes (containing the main categories social processes, affective processes, cognitive processes, perceptual processes, biological processes, and relativity), personal concerns, and spoken language (Pennebaker, et al., 2007). Each of these categories has some features. For example, the category "social words" contains the features "family", "friends", and "humans", and the category "relativity" is divided into the features "motion", "space" and "time". We use LIWC2007 to extract these features from the motivation statements and to use as input for the models. In this version of LIWC, its dictionaries have been translated into several languages, including Dutch (Boot et al., 2017). Together with the 15 topics, we refer to the LIWC sub-categories as "features extracted from text" in the remainder of this paper.

*2.2.3. Training the algorithm*
Support Vector Machine (SVM) was chosen to analyze the data, since it can handle (the combination of) structured and unstructured data, and is known to generally perform well in text classification problems. First, the data were randomly split into training (75%; N = 5295)) and test sets (25%; N = 1765). The training set was used to train the algorithm by providing it with both the input (i.e., student characteristics, text, and text features) and the output (i.e., whether a student dropped out or not). K-fold cross validation with k = 5 was used to evaluate the performance of the training model (Refaeilzadeh et al., 2009). Cross validation is a resampling process in which the training data is split in k-different portions to test and train a model on different iterations. If the different iterations return different levels of model accuracy, this can indicate potential problems regarding overfitting or selection bias. The full set of training data can then be inspected further to ensure better performance of the algorithm when eventually applied to the test data. For none of the estimated models in our study cross validation indicated a need for further inspection of the training data.

*2.2.4. Analysis*
We analyzed our data using six separate SVM models, exploring the most accurate combination of features to predict dropout. First, we started with a model using only the structured data, our set of student characteristics, as input for the model. This provided us with a baseline of how well we would be able to predict dropout if we would not have any text data. Second, we estimated a model with only the text using TFiDF as input to compare the algorithms' performance to that of the first model. Third, we added the features that we extracted from the text through LDA topic modeling and the LIWC dictionary to the text-only model to assess the added value of more advanced text mining techniques on top of the simple bag-of-words representation. Lastly, to answer the question whether information extracted from text can add to the prediction of dropout net of structured data, we examined the performance of the SVM with different combined feature sets. Table 2 provides an overview of the input per model.

**Table 2.** Overview of the feature sets across the different models.

| Features | Model 1 | Model 2 | Model 3 | Model 4 | Model 5 | Model 6 |
|---|---|---|---|---|---|---|
| **Structured data** | | | | | | |
| Prior education | X | | | X | X | X |
| High school performance | X | | | X | X | X |
| Ability beliefs | X | | | X | X | X |
| Interests | X | | | X | X | X |
| Gender | X | | | X | X | X |
| Age | X | | | X | X | X |
| Cohort | X | | | X | X | X |
| Discipline | X | | | X | X | X |
| Study program | X | | | X | X | X |
| Previously enrolled | X | | | X | X | X |
| Multiple requests | X | | | X | X | X |
| **Unstructured data** | | | | | | |
| Text (TFiDF) | | X | X | X | | X |
| Latent Dirichlet Allocation | | | X | | X | X |
| LIWC dictionary | | | X | | X | X |

Student retention data are generally imbalanced, since the number of dropouts is much smaller than the number of students that continue the study program. This imbalance can be problematic, as the standard classification algorithms have a bias towards the majority class, giving misleadingly promising results (Dalipi et al., 2018). To solve the issue of imbalanced data, we applied a balanced class weight with all our classification algorithms (see GitHub for code). This balanced weight can be considered an oversampling technique which essentially replicates the smaller class until it reaches the sample size of the larger one.

Performance of Support Vector Machines is generally assessed by accuracy, precision, recall, and f1-scores. Accuracy is an unreliable measure for imbalanced data and therefore we do not use it. Moreover, because of the imbalance, weighted output for precision, recall, and f1-score are reported to assess the performance of the algorithm. Precision denotes the true positives divided by the true positives + false positives. Recall is defined as the true positives divided by the true positives + false negatives. The f1-score is the weighted average of precision and recall. To get a better sense of these performance measures for our specific context, Figure 1 shows precision and recall for the dropout class.

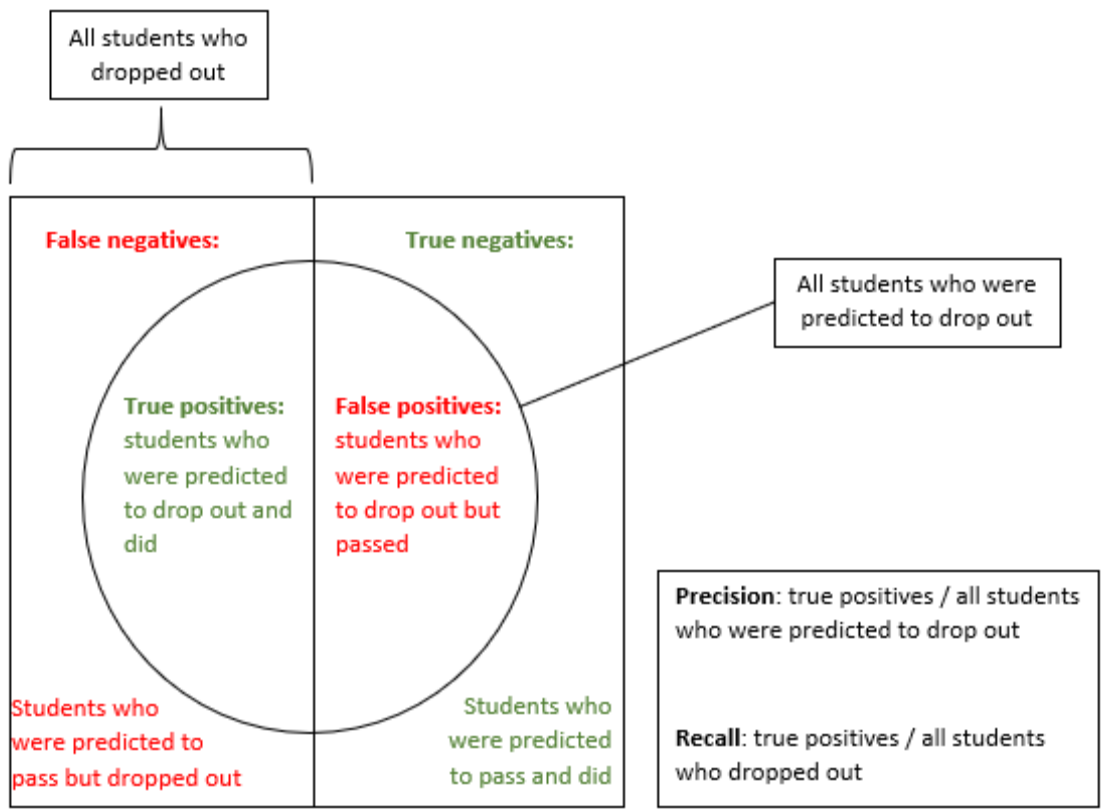

**Figure 1.** Visualization of the model performance criteria precision and recall, applied to the dropout class.

# 3. Results

The total dataset was split into a training set (75%) and a test set (25%). Results presented in this section are based on the 1765 motivation statements forming the test data. Of these 1765 statements, 1312 belonged to the retention class and 453 to the dropout class. Table 3 provides a summary of the results. Given our dual aim of exploring whether students at risk of dropout can be identified through motivation statements and whether information extracted from these motivation statements adds predictive power net of student characteristics, we are interested in model performances in absolute terms, but not in whether certain models statistically outperform others.

**Table 3.** Performance measures of the estimated models, for the total test set (T) and split by the retention (R) and dropout (D) class.

|       | Precision |     |     | Recall |     |     | f1-score |     |     |
|-------|-----------|-----|-----|--------|-----|-----|----------|-----|-----|
| Model | T         | R   | D   | T      | R   | D   | T        | R   | D   |
| 1     | .71       | .84 | .31 | .57    | .55 | .65 | .60      | .66 | .42 |
| 2     | .67       | .79 | .28 | .63    | .71 | .37 | .65      | .75 | .32 |
| 3     | .67       | .79 | .29 | .64    | .72 | .37 | .65      | .75 | .32 |
| 4     | .69       | .81 | .32 | .65    | .71 | .44 | .67      | .76 | .37 |
| 5     | .70       | .82 | .31 | .60    | .61 | .56 | .63      | .70 | .40 |
| 6     | .68       | .80 | .31 | .64    | .71 | .42 | .66      | .75 | .35 |

## 3.1. Model 1: Student characteristics

First, a SVM model with only student characteristics was run as a baseline model. This model contains variables that are known to be predictive of student dropout, including high school performance. The weighted precision score of Model 1 was .71, the weighted recall score .57. This resulted in a weighted f1-score of .60. The prediction of precision for the majority class (retention) was much better than the performance for the minority class (dropout). With a score of .65, recall was higher for the dropout class than for the retention class (.55). This is notable, given the fact that algorithms generally perform better for the majority class, even after correcting for imbalanced data.

## 3.2. Model 2: Only text

To identify what the analysis of text data through NLP-based techniques can add to the prediction of dropout, first a model with only text was analyzed, using TFiDF to identify the importance of words in the corpus (Model 2). The model had a weighted precision score of .67, a weighted recall score of .63, and a weighted f1-score of .65. When comparing the performance of the algorithm for this model several things stand out. First, precision of this model is worse than in Model 1, meaning that in the model with the student characteristics, of all the selected students there were proportionally more true positives. In Model 2 recall is better than in Model 1, meaning that of all the relevant students there were proportionally more true positives in the model with only text. Second, recall scores for this model are less balanced across the classes than in Model 1. Table 3 shows a much higher recall score for the retention class (.71) than for the dropout class (.37).

## 3.3. Model 3: text and text features

Model 3 investigates the predictive power of all information we extracted from text. To that end, we added features extracted from text to the text data (LIWC dictionary words and topics extracted through LDA topic modelling), to investigate whether this combination could outperform Model 1. The LIWC features already consist of predetermined categories, using psychometrically validated dictionaries. For the LDA topic modelling the features first must be identified before they can be used in the analyses. For the 15 topics that were identified in the motivation statements, the top 10 terms of each of these topics are listed in Table 4. Upon consensus among the authors, the 15 topics were given a theoretical label if possible. Some of the topics did not construe one clear underlying concept and were therefore left unlabeled.

**Table 4.** Top ten most used terms per topic, resulting from the LDA topic modelling.

| Topic | Label | Top words |
|---|---|---|
| 1 | program interest | study, Utrecht, program, very, finds, fun, seems, good, very, rather |
| 2 | general interest | university, highly, knowledge, Utrecht, study, interest, offers, like, choice, developing |
| 3 | previously enrolled | year, study, go, wanted, came, found, rather, choice, studying, knew |
| 4 | applied sciences | program, university of applied sciences, Utrecht, scientific education, university, less, aspects, academic, difference, choose |
| 5 | culture minded | subjects, different, language, interests, culture, year, broad, cultures, choosing, liberal arts |
| 6 | sense of belonging | day, open, trial studying days, visited, found, open days, during, ambiance, spoke, immediately |
| 7 | societal minded | social, spatial planning, study, sciences, general, geography, studies, different, expect, hope |
| 8 | pedagogical minded | children, study, pedagogical sciences, chosen, helping, doubts, good, characteristics, later, finished |
| 9 | computer minded | programming, games, computers, artificial intelligence, game technology, suitable, technology, game, computer, logic |
| 10 | artistic students | media, art, theater, culture, film, television, films, chose, theater film, Amsterdam |
| 11 | - | music, nature, hopefully, astronomy, most important, therein, teaching school, ever, madam, dear sir |
| 12 | - | person, maximum, character, getting acquainted, function, fascinated, legal system, honest, nature (kind), wonderful |
| 13 | location | location, widening, exist, per, passed, stadium, analysis, classes, acquaintances, about |
| 14 | politics minded | political, strike, stone (figure of speech), technological, horizon, sustainable, advance, curriculum |
| 15 | - | help, automatically, job opportunity, sociological, public, mono disciplinary, suits |

Figure 2 shows the different topics ordered by strength for predicting dropout (positive values) and retention (negative values). The topics 13 (location), 7 (societal minded), and 11 (unlabeled) were the strongest positive predictors of dropout. Students using a lot of words related to these aspects, are more likely to drop out. The strongest negative predictors of dropout were the topics 2 (program interest), 12 (unlabeled), and 1 (general interest). Students using many of such words, are less likely to drop out.

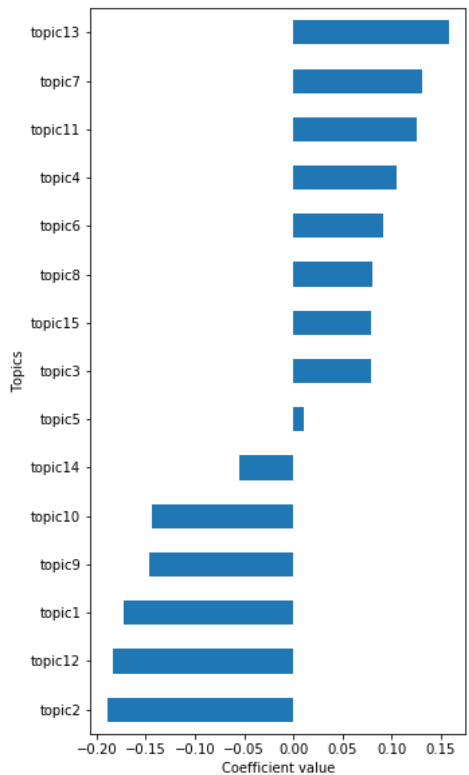

**Figure 2.** Topics extracted through LDA topic modelling, ordered by strength (x-axis coefficient) for predicting dropout (positive values) and retention (negative values).

Subsequently, all 15 topics were fed to the SVM, together with text data of Model 2 and features from the LIWC package for text analysis. The results in Table 3 show that Model 3 is almost identical to Model 2. In other words, it appears that, in this study, the features extracted through LDA topic modelling and the LIWC package do not add to the prediction of dropout in comparison to text alone.

### 3.4. Model 4: Student characteristics and text

To identify whether written motivation adds to the prediction of first-year dropout net of student characteristics, Model 1 was combined with different elements of text. In Model 4 the input of Model 1 (student characteristics) and Model 2 (text only) was combined. The weighted recall (.65) and weighted f1-score (.67) in this model are the highest of all the estimated models. The weighted precision score (.69) of this Model holds a middle position regarding algorithm performance between Model 2 and 3 on the one hand, and Model 1 on the other hand.

## 3.5. Model 5: Student characteristics and text features

In this fifth model, student characteristics were combined with features extracted from the text, rather than the text itself. Even though the features extracted from text did not add to the predictive power of dropout net of text alone (Model 3), the combination of student characteristics and features extracted from text might improve the prediction of dropout. With a weighted precision score of .70, a weighted recall score of .60, and a weighted f1-score of .63, this algorithm performed worse than in Model 4.

Model 5 was also used to inspect the importance of all features together that were used as input for the SVM (i.e., student characteristics, LIWC words, and LDA topics). We performed this step on this model, rather than Model 6 (see below), because the vast number of features of the TFiDF method (i.e., all the individual words in the texts) does not allow it to be captured in a figure. The strength of the 25 most important features for predicting dropout (positive values) and retention (negative values) is shown in Figure 3. Some of the most important features are discussed. The strongest positive effect is word count (WC), indicating that the more words students used the higher their probability of dropout. Second, the use of personal pronouns (ppron) is a similarly strong predictor of dropout in this model. The more personal pronouns a student uses in their text, the higher the probability of dropout. Frequent use of the first person singular (i), however, is negatively related to dropout. Looking at Figure 3, it indeed seems to matter which pronouns are being used. For example, the more the second person (you) is used, the lower the probability of dropout, whereas the relatively frequent use of impersonal pronouns (ipron) is associated with a higher probability of dropout. It is noteworthy that, in this model, high school performance was again the strongest negative predictor of dropout. Lastly, relatively important features in this list are age and article. Age was a relatively weak predictor in the model with only student characteristics. In this model, however, it is the third most important predictor of dropout, with older students having a higher probability to drop out. The relatively frequent use of articles (article), on the other hand, is associated with a lower probability to drop out. Among the top 25 most important features for the prediction of dropout there are several other features that were obtained through the LIWC dictionary. Interestingly though, none of the topics from the LDA topic modelling is amongst the 25 most important predictors of dropout.

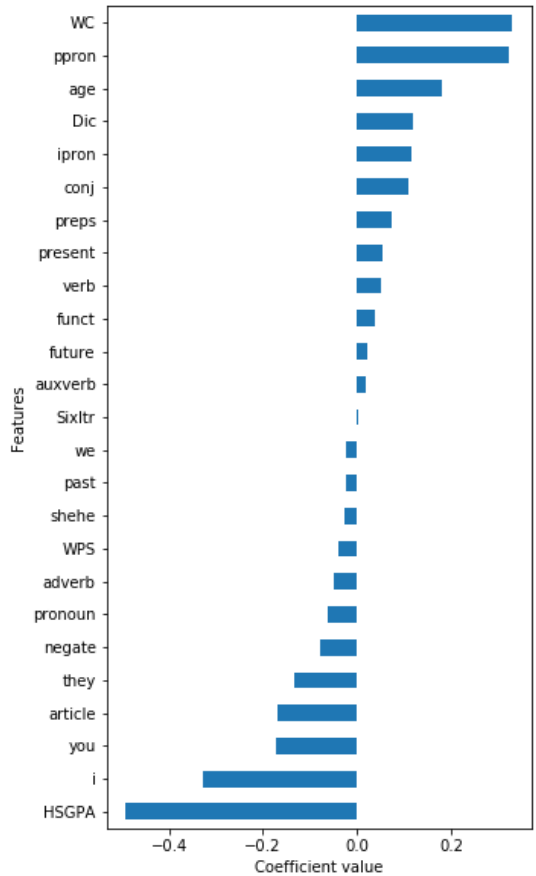

**Figure 3.** 25 most predictive features[3] of Model 5, ordered by strength (x-axis coefficient) for predicting dropout (positive values) and retention (negative values).

### 3.6. Model 6: Student characteristics, text, and text features

Lastly, a model was estimated that included the student characteristics, as well as the text itself and the features extracted from the text. The algorithm performance of this model was almost the same as the performance for Model 4. The weighted precision was .68, weighted recall .64, and the weighted f1-score .66. When comparing this model to Model 5, it strengthens the conclusion that features extracted from text are, in our dataset, of limited additional value in the prediction of dropout.

---

[3] Meaning of the features from top to down: word count (WC); personal pronouns (ppron); age; dictionary words (Dic); impersonal pronous (ipron); conjunctions (conj); prepositions (preps); present tense (present); common verbs (verb); total function words (funct); future tense (future); auxiliary verbs (auxverb); words with more than 6 letters (sixltr); first person plural (we); past tense (past); third person singular (shehe); words per sentence (WPS); adverbs (adverb); total pronouns (pronoun); negations (negate); third person plural (they); articles (article); second person (you); first person singular (i); HSGPA (high school performance).

## 3.7. Comparing most frequently used terms of correctly and incorrectly classified dropouts

To get a better understanding of the algorithm performance, top terms used by students who were correctly identified as dropouts, were compared to top terms used by students who were incorrectly classified as dropouts. A high overlap in the commonly used terms, would indicate that there is not enough discrepancy in the written motivation between the two groups for the SVM to detect any differences.

When we inspected the 100 most used terms for both groups, overlap was indeed identified. Roughly a quarter of the top terms was used by students in both categories. Most of these words were names of study programs (e.g., *Biology, Law, Sociology*), or derivatives thereof (e.g., *game(s)* for Game Technology or *art* for Art History). The other overlapping words are generic, such as *program* or *name*, or apply to a specific field (i.e., *people* and *children* for behavioral sciences and *music* and *film* for arts programs). Given that most of the overlapping words refer to names of study programs or derivatives thereof, the prediction of dropout may improve if these words can be excluded from the text input. Because of the too small sample size per study program in our data we were not able to do this.

## 4. Discussion

In this study we attempted to answer the question whether students at risk of dropout can be identified based on their motivation for the study program of their initial choice as written in their intake questionnaire prior to enrollment by using NLP-based techniques. Moreover, we asked the question whether information extracted from these motivation statements adds predictive power net of student characteristics. The results showed that the answer to this question is twofold. When text was used in addition to student characteristics, it hardly added to the prediction of dropout. However, when only text data were used, the algorithm performed very similar to the one in the model with only the student characteristics.

Predicting dropout accurately is not easy, especially not based on student information that is available prior to enrolment in higher education. Since the model with only text showed very similar results to the model with only student characteristics, it appears that student dropout can be predicted with a short motivation statement analyzed with data mining techniques at least as good as with a set of known predictors like high school performance. Once it becomes known exactly what aspects of motivation texts are predictive of dropout, these features might be easy to manipulate. Therefore, future research should focus on the reliability of motivation texts in comparison to student characteristics.

Moreover, structured and unstructured data seem to complement each other, as precision was higher in the model with student characteristics (Model 1) and recall was higher in the model with only text (Model 2). Therefore, analyzing text data with text mining techniques seems promising. Our aim was to exhume hidden information from text data and investigate whether this information could be used to predict students at risk of dropout. Unstructured data, like text, are very time consuming and complex to analyze for humans. However, if highly predictive text mining algorithms can be developed to analyze these data, that could potentially be useful to identify students at risk of dropout before the start of the study program without needing an extensive set of student characteristics. Such students could then immediately be offered support to mitigate the risk.

The fact that combining the text and student characteristics, like high school performance, does not (substantially) improve the model prediction in this study, might indicate that they measure the same underlying concepts. It is possible that the way the question about motivation for the study program was asked or explained, probes students to put into words the information they already filled out earlier in the questionnaire by answering the other survey questions. Future research could try to verify this hypothesis by studying motivation statements using less directive questions. Another possible way might be to ask more open-ended questions about specific components of program choice (e.g., why this study program; why this university; what in this study program makes it attractive; etc.) to obtain more unstructured (i.e., text) data covering a variety of underlying concepts to analyze and relate to academic outcomes, using machine learning techniques. It was beyond the scope of this paper to compare the performance of the SVM to other machine learning techniques, like K-nearest neighbors or naïve Bayes (e.g., Aggarwal et al., 2012, for comparing the performance of different techniques), but such an approach could be considered in the future.

A limitation of this study lies in the properties of our dataset. First, the dataset is imbalanced because there are proportionally few dropouts. This is generally the case with student retention data, and therefore, cannot be solved. Oversampling and undersampling techniques were used and weighted scores were reported to deal with this limitation. Second, the motivation statements are generally short (i.e., students were requested to write 10-25 lines, resulting in texts that are roughly 250 words long) and the dataset consists of applicants to all non-selective bachelor programs. Both the length of the texts and the heterogeneous student population may have an influence on the ability of the algorithm to construct an accurate prediction model. Algorithms learn by providing them with more and more data. Despite our relatively large number of motivation statements, the relatively short and pluriform texts that were used could have affected the performance of the algorithm for the text models. Future research may investigate whether a more uniform group of students (e.g., of one faculty or one study program) would result in a better performance of the text mining approach.

Another direction for future research is to apply deep learning-based NLP methods with the use of transfer learning (i.e., improving learning in a new task through the transfer of knowledge acquired in a previous task) on a bigger dataset. This could improve representation of text data using the distributional hypothesis, which poses that the more semantically similar two words are, the more distributionally similar they will be, and thus the more they tend to occur in similar linguistic contexts (Sahlgren, 2008). For example, the algorithm can use the fact that the words *city* and *location* are distributionally more like one another than they are to the word *scientific* in a multidimensional context, to predict that *city* and *location* are also semantically more like one another than to the word *scientific*. However, these techniques require more data. Nevertheless, this direction is worth researching as it could help in capturing the distinctive writing style of a student. This information could, in turn, contribute to the early identification of dropouts.

When developing prediction models with the aim to use them for the early identification of dropouts, one should especially focus on improving the precision scores for the dropout class. If text mining methods were to become reliable enough to be used in advising students on their program choice, we think it would be better to incorrectly classify a student as successful, than to incorrectly classify a student as a future dropout. Some students who would then be incorrectly advised to start the study program might actually be successful if the positive advice has an influence on their feelings of self-efficacy and/or motivation. Regardless of whether an advice can have such an effect, we think it is better to have a predictive instrument that returns very few false positives (students unjustly classified as dropout) than an instrument that returns very few false negatives (students who were unjustly classified as successful). Therefore, if choices must be made in developing these models, prioritizing a high precision score for the dropout category is most favorable when one wants to give every student a chance.

There is a famous quote by the economist Ronald Coase stating: *"if you torture the data long enough, it will confess to anything".* Coase meant this as a warning to fellow researchers, not to engage in scientific misconduct (e.g., p-hacking). Although not for the purposes of making it confess to anything specific, torturing the data is exactly what we did in this study. We approached the motivation data from different angles, to predict first-year dropout of students applying for a certain undergraduate program. By comparing and combining different methods, we found that applying machine learning techniques on student motivation is a potentially promising new way of approaching student dropout. We believe it is worthwhile to explore this line of research further to find better ways to extract information from motivation statements. When doing so, focus could also be placed on comparing human-coded and algorithm-coded motivation statements to get a better sense of how accurate these methods are in predicting dropout and which of them is better.